\documentclass[twocolumn,twoside,preprintnumbers,amsmath,amssymb,pacs]{revtex4}
\usepackage{epsfig}
\usepackage{graphicx}% Include figure files
\usepackage{dcolumn}% Align table columns on decimal point
\usepackage{bm}% bold math

\usepackage{fancyhdr}

\usepackage{pslatex}

\begin{document}

\title{{\Large Condensation of Vortex-Strings: Effective Potential
    Contribution Through Dual Actions }}

\author{Rudnei O. Ramos} \email{rudnei@uerj.br} \affiliation{Departamento de
  F\'{\i}sica Te\'orica, Universidade do Estado do Rio de Janeiro, 20550-013
  Rio de Janeiro, RJ, Brazil}

\author{Daniel G. Barci} \email{barci@uerj.br} \affiliation{Departamento de
  F\'{\i}sica Te\'orica, Universidade do Estado do Rio de Janeiro, 20550-013
  Rio de Janeiro, RJ, Brazil}

\author{Cesar A. Linhares} \email{linhares@dft.if.uerj.br}
\affiliation{Departamento de F\'{\i}sica Te\'orica, Universidade do Estado do
  Rio de Janeiro, 20550-013 Rio de Janeiro, RJ, Brazil}

\author{J. F. Medeiros Neto} \email{jfmn@ufpa.br} \affiliation{Instituto de
  F\'{\i}sica, Universidade Federal do Par\'a, 66075-110 Belem, Par\'a,
  Brazil}

%\author{}

%\affiliation{ }

%\affiliation{ }

%\received{on 24 March, 2006}

\begin{abstract}
  
  Topological excitations are believed to play an important role in different
  areas of physics. For example, one case of topical interest is the use of
  dual models of quantum cromodynamics to understand properties of its vacuum
  and confinement through the condensation of magnetic monopoles and vortices.
  Other applications are related to the role of these topological excitations,
  nonhomogeneous solutions of the field equations, in phase transitions
  associated to spontaneous symmetry breaking in gauge theories, whose study
  is of importance in phase transitions in the early universe, for instance.
  Here we show a derivation of a model dual to the scalar Abelian Higgs model
  where its topological excitations, namely vortex-strings, become manifest
  and can be treated in a quantum field theory way. The derivation of the
  nontrivial contribution of these vacuum excitations to phase transitions and
  its analogy with superconductivity is then made possible and they are
  studied here.
  
  

\end{abstract}

\maketitle

\thispagestyle{fancy} %\setcounter{page}{0}

\section{Introduction}

Topological excitations, or defects, are nonhomogeneous solutions of the field
equations of motion in many types of field theory models
\cite{coleman,rajaraman,review}.  They are finite energy and stable
configurations that emerge as a consequence of a spontaneous symmetry breaking
process. Mathematically, defects are predicted to appear whenever some larger
group of symmetry $G$ breaks into a smaller one $H$ such that there are a
nontrivial homotopy group $\pi_k(G/H)$ of the vacuum manifold different from
the identity.  Well known examples are kinks, or domain walls ($k=0$)
\cite{rajaraman}, that originate from a discrete symmetry breaking, strings,
or vortices ($k=1$), for example originating from a continuous gauge symmetry
breaking $U(1) \to 1$ \cite{olesen} and magnetic monopoles ($k=2$), e.g. from
a $SO(3) \to U(1)$ symmetry breaking \cite{hooft,polyakov}.  Since many phase
transitions in nature are associated to symmetry breakings, topological
excitations are a common feature in these processes and are in fact observed
in many systems in the laboratory, like in ferromagnetism, helium
superfluidity, superconductivity and in many other condensed matter system and
they are also expected to have appeared in phase transitions in the early
universe as well (for a general review, please see \cite{kibble}).

In the study of phase transitions in quantum field theory one basic quantity
usually computed is the effective potential, which is an important tool in the
study of phase transitions in scalar and gauge field theories \cite{effpot}.
It is equivalent to the homogeneous coarse-grained free-energy density
functional of statistical physics, with its minima giving the stable and, when
applicable, metastable states of the system.  {}For interacting field theories
the effective potential is evaluated perturbatively, with an expansion in
loops being equivalent to an expansion in powers of $\hbar$ \cite{CW}. The
one-loop approximation is then equivalent to incorporating the first quantum
corrections to the classical potential. Recall that the effective potential,
taking a scalar field theory as an example, is obtained from the effective
action $\Gamma[\phi_c]$, where it is defined in terms of the connected
generating functional $W[J]$ as

\begin{equation}
\Gamma[\phi_c]=W[J]-\int d^4x J(x)\phi_c(x) \:~~,
\label{eq:2}
\end{equation}
with the classical field $\phi_c(\vec{x},t)$ defined by

$\phi_c(\vec{x},t)\equiv \delta W[J]/\delta J(x)$, and

\begin{equation}
W[J]=-i\hbar {\rm ln}\int D \phi ~ {\rm exp}\left [\frac{i}{\hbar}S[\phi,J]
\right ] \:.
\label{e:3}
\end{equation}

In order to evaluate $\Gamma[\phi_c]$ perturbatively, one writes the field as
$\phi(\vec{x},t) \rightarrow \phi_{0}(\vec{x},t) + \eta(\vec{x},t)$, where
$\phi_{0}(\vec{x},t)$ is a field configuration which extremizes the classical
action $S[\phi,J]$, $\frac{\delta S[\phi,J]}{\delta \phi} |_{\phi = \phi_{0}}
= 0$, and $\eta ({\vec{x},t})$ is a small perturbation about that extremum
configuration.  The action $S[\phi,J]$ can then be expanded about
$\phi_0(\vec{x},t)$ and, up to quadratic order in $\eta(\vec{x},t)$, we can
use a saddle-point approximation to the path integral to obtain for the
connected generating functional,

\begin{equation}
W[J]=S[\phi_0]+\hbar\int d^4x \phi_0(x)J(x)+i\frac{\hbar}{2}
{\rm Tr ln}\left [\partial_{\mu}\partial^{\mu}+V^{\prime \prime}
(\phi_0)\right ] \:.
\label{eq:4}
\end{equation}
In order to obtain the one-loop expression for $\Gamma[\phi_c]$, we first note
that writing $\phi_0=\phi_c-\eta$ we get to first order in $\hbar$,
$S[\phi_0]=S[\phi_c]-\hbar\int d^4x \eta(x)J(x)+{\cal O}(\hbar^2)$.  Using
this result and Eq. (\ref{eq:4}) into Eq. (\ref{eq:2}) we find, as
$J\rightarrow 0$,

\begin{equation}
\Gamma[\phi_c]=S[\phi_c]+i\frac{\hbar}{2}{\rm Tr ln}
\left [\partial_{\mu}\partial^{\mu}+V^{\prime \prime}
(\phi_c)\right ] \:.
\label{eq:5}
\end{equation}
The effective action can also be computed as a derivative expansion about
$\phi_c(\vec{x},t)$,

\begin{equation}
\Gamma[\phi_c]=\int d^4x\left [-V_{\rm eff}(\phi_c(x))+\frac{1}{2}
\left (\partial_{\mu}
\phi_c\right )^2Z(\phi_c(x))+\dots \right ] \:.
\label{e:6}
\end{equation}
The function $V_{\rm eff}(\phi_c)$ is the effective potential. {}For a
constant field configuration $\phi_c(\vec{x},t)=\phi_c$ we obtain

\begin{equation}
\Gamma[\phi_c]=-\Omega V_{\rm eff}(\phi_c) \:,
\label{eq:7}
\end{equation}
where $\Omega$ is the total volume of space-time. Comparing Eqs. (\ref{eq:5})
and (\ref{eq:7}) we obtain for the one-loop effective potential,

\begin{equation}
V_{\rm eff}(\phi_c)=V(\phi_c)-i\frac{\hbar}{2}\Omega^{-1}{\rm Tr ln}
\left [\partial_{\mu}\partial^{\mu}+V^{\prime \prime}
(\phi_c)\right ] \:.
\label{eq:8}
\end{equation} 

When working at non-vanishing temperature, the same functional techniques can
be used. In this case one is interested in evaluating the generating
functional (the partition function) $Z_{\beta}[J]$ which is given by the path
integral \cite{kapusta}

\begin{equation}
Z_{\beta}[J]=N\int D\phi {\rm exp}\left [-\int_0^{\beta}d\tau \int d^3x\left
({\cal L}_E-J\phi\right )\right ] \:,
\label{eq:9}
\end{equation}
where the integration is restricted to paths periodic in $\tau$ with
$\phi(0,\vec{x})=\phi(\beta,\vec{x})$, ${\cal L}_E$ is the Euclidean
Lagrangian, and $N$ is a normalization constant. Again one expands about an
extremum of the Euclidean action and calculates the partition function by a
saddle-point evaluation of the path integral.  The result for the one-loop
approximation to the effective potential is

%\begin{widetext}
%\begin{equation}
%\end{equation}
%\end{widetext}
\begin{eqnarray}
\lefteqn{V_{\rm eff}(\phi_c,T) = V_{\rm eff}(\phi_c)} \nonumber \\
&&+ \frac{\hbar}{2\pi^2\beta^4}
\int_0^{\infty}dx~x^2{\rm ln}\left \{1-{\rm exp}\left [-\sqrt{x^2+\beta^2
V^{\prime \prime}(\phi_c)}\right ] \right \} \:~.
\label{eq:10}
\end{eqnarray}
As the effective potential is equivalent to the free energy functional (for a
constant field configuration), all thermodynamics functions follow from it.
In particular the different phases, critical temperature of phase transition
and temperature dependence of the field vacuum expectation value can be
obtained from (\ref{eq:10}).

{}From the above discussion it is clear that the one-loop approximation to the
effective action, Eq. (\ref{eq:5}), works best when the classical field does
not differ much from the configuration that extremizes the classical action,
$\phi_c=\phi_0 +\eta\sim \phi_0$, since in this case the saddle-point
evaluation to the path integral is adequate.  Also, $\phi_c(\vec{x},t)$ must
be nearly constant so that the effective potential can be obtained from Eq.
(\ref{eq:7}). As $J\rightarrow 0,~\phi_c(\vec{x},t)$ is identified with
$\langle\phi\rangle$, the vacuum expectation value. One major problem we see
in this whole approach of studying the phase structure of field theory models
from the effective potential is when the action functional, determined from
Eq. (\ref{eq:9}) is dominated not by homogeneous, constant field
configurations but by nonhomogeneous ones. In those situations when other
stable, finite energy solutions to the field equations of motion exist, we
expect these configurations to dominate the partition function over the
homogenous solutions for instance close to the critical temperature
\cite{freeenergy}. Under these circumstances the effective potential, which
includes only contributions to the partition function from constant background
field configurations becomes inappropriate to study the phase transition and we
must rely in other approaches, for example studying the phase transition
directly from the effective action or free energy for the topological
configurations \cite{freeenergy,copeland}, or taking directly a field
theoretic description for the topological excitations \cite{MMR}. In either
case we are faced with the problem of accounting for nonlocal contributions in
the perturbative expansion, which is only amenable of analysis up to the
lowest order leading order.  To circumvent these difficulties we make use of
the techniques of dualization in field theory, from which the degrees of
freedom of the topological excitations are explicitly realized in the
functional action. This method is described in the following section, Sec.
\ref{sec2}, where we specialize to the case of the scalar Abelian Higgs
model, whose topological solutions are vortex-strings. 
In Sec. \ref{dualaction} we evaluated the dual action for the model,
making explicit the vortex-strings degrees of freedom and how they couple
to the matter fields. In Sec. \ref{sec3} we
show how an effective potential calculation for an averaged vortex-string
field can be computed and the results and interpretation of the phase
transition obtained from this quantity. {}Finally, we give our conclusions in
Sec. \ref{conclusions}.

\section{The String solutions in the Scalar Abelian Higgs Model}
\label{sec2}

In this work we will use the Abelian Higgs model, with Lagrangian density for
a complex scalar field $\phi$ and gauge field $A_\mu$ given by

\begin{equation}
{\cal L} = - \frac{1}{4} F_{\mu \nu } F^{\mu \nu }+|D_\mu \phi|^2
-V(\phi)\;,
\label{lagr}
\end{equation}

\noindent
where, ${}F_{\mu \nu } =\partial _\mu A_\nu -\partial _\nu A_\mu$, $D_\mu =
\partial _\mu -ieA_\mu$ and $V(\phi)$ is a symmetry breaking potential, for
example given by

\begin{equation}
V(\phi) = - m_{\phi }^2\left| \phi \right| ^2+\frac \lambda
{3!}\left( \left| \phi \right| ^2\right) ^2 \;,
\label{pot}
\end{equation}

\noindent
with positive parameters $m_\phi^2$ and $\lambda$.  The symmetry breaking
$U(1) \to 1$ with homotopy group $\pi_1 \neq 1$ indicates the existence of
string-like topological excitations in the system, or Nielsen-Olesen strings
\cite{olesen} (for an extended introduction and review see e.g. Ref.
\cite{review}). {}For example, for a unit winding string solution along the
$z$ axis, the classical field equations of motion obtained from the Lagrangian
density (\ref{lagr}) admit a stable finite energy configuration describing the
string given by (using the cylindrical coordinates $r,\theta,z$)

\begin{eqnarray}
\phi_{\rm string} &=& 
\frac{\rho(r)}{\sqrt{2}} e^{i \theta}\;,
\label{phi string}
\\
A_{\mu, {\rm string}}  &=& \frac{1}{e} A(r)\; \partial_\mu \theta\;,
\label{A string}
\end{eqnarray}

\noindent
where the functions $\rho(r)$ and $A(r)$ vanish at the origin and have the
asymptotic behavior

\begin{eqnarray}
\phi(r \to \infty) &\to & \rho_v \equiv \sqrt{\frac{6 m_\phi^2}{\lambda}}\;,
\nonumber \\ 
A(r \to \infty) &\to & 1\;.
\end{eqnarray}

\noindent
The functions $\rho(r)$ and $A(r)$ can be obtained numerically by solving the
classical field equations for $\phi$ and $A_\mu$.  If we write the field
$\phi$ as $\phi =\rho \exp (i\chi )/\sqrt{2}$, then from (\ref{phi string})
and (\ref{A string}) for the string, at spatial infinity $\rho$ goes to the
vacuum $\rho_v$ and $A_\mu$ becomes a pure gauge. This also gives, in order to
get a finite energy for the string configuration, that $\partial_\mu \chi = e
A_\mu $ at $r\to \infty$, so $D_\mu \phi=0$. This leads then that, by taking
some contour $C$ surrounding the symmetry axis, and using Stokes' theorem, to
the nonvanishing magnetic flux

\begin{equation}
\Phi = \oint A_\mu dx^\mu = \oint \partial_\mu \chi dx^\mu = 2\pi/e\;.
\label{flux}
\end{equation}

\noindent
Since $\phi$ must be single-valued, the Eq. (\ref{flux}) implies that on the
string $\chi$ must be singular. Therefore, the phase $\chi$ can be separated
into two parts: in a regular part and in a singular one, due to the string
configuration,

\begin{equation}
\chi(x) = \chi_{\rm reg}(x) + \chi_{\rm sing}(x)\;,
\label{chiparts}
\end{equation}

\noindent
where the singular (multivalued) part $\chi_{\rm sing}(x)$ can be related to a
closed world-sheet of an vortex-string \cite{dirac},

\begin{eqnarray}
\frac{1}{2\pi }\epsilon_{\mu \nu \lambda \rho} 
\partial_\lambda \partial_\rho
\chi_{\rm sing} (x)=n\int_Sd\sigma_{\mu \nu }(x)\delta^4 [x-y(\xi )] =
\omega_{\mu \nu} \;,
\label{vort2}
\end{eqnarray}

\noindent
where $n$ is a topological quantum number, the winding number, which we here
restrict to the lowest values, $n=\pm 1$, corresponding to the energetically
dominant configurations. The element of area on the world sheet swept by the
string is given by

\begin{eqnarray}
d\sigma_{\mu \nu }(x)=\left( \frac{\partial x_\mu }{\partial \xi ^0}
\frac{\partial x_\nu }{\partial \xi^1}-\frac{\partial x_\mu }{\partial \xi^1}
\frac{\partial x_\nu }{\partial \xi^0}\right) d^2\xi
\label{sigma}
\end{eqnarray}

\noindent
and $y_\mu (\xi )$ represents a point on the world sheet $S$ of the
vortex-string, with internal coordinates $\xi^0$ and $\xi^1$. As usual, we
consider that $\xi^1$ is a periodic variable, since we work with closed
strings, whereas $\xi^0$ will be proportional to the time variable (at zero
temperature), in such a way that $\xi^1$ parameterizes a closed string at a
given instant $\xi^0$. Eq. (\ref{vort2}) is known as the vorticity.  Eqs.
(\ref{vort2}) and (\ref{sigma}) entails the vortex-string degrees of freedom
and then can be used to identify the topological vortex string contributions
to the partition function.

Let us briefly recall two main previous methods to take into account the
effect of topological strings in phase transitions. The first attempt to do so
made use of semiclassical methods \cite{rajaraman}. In the semiclassical
method we use directly the nonhomogeneous string solutions, Eqs. (\ref{phi
  string}) and (\ref{A string}), when evaluating the effective action. In this
case the effective action is evaluated after taking fluctuations around the
string vacuum solutions, $\phi \to \phi_{\rm string} + \phi'$ and $A_\mu \to
A_{\mu, {\rm string}} + A_\mu'$ and the functional integration performed over
the fluctuation fields $\phi'$ and $A_\mu'$. In the one-loop approximation,
this gives the analogous to Eq. (\ref{eq:5}), with the constant background
field now replaced by the scalar string background configuration plus those
analogous loop corrections for the gauge field string configuration.  But from
Eqs. (\ref{eq:5}) and (\ref{eq:9}), we see that the effective action at finite
temperature is just associated with the free energy of the system, where, here
is the free energy in the presence of the string field configurations.  This
is the procedure used for instance in the papers in Ref. \cite{freeenergy}.
The free energy relevant for the study is written as \cite{freeenergy}

\begin{equation}
{}F_{\rm string} = - \frac{1}{\beta L} {\rm ln} \left(
\frac{Z_{\rm string}}{Z_v} \right)\;,
\label{Fs}
\end{equation}

\noindent
where $Z_{\rm string}$ is the partition function evaluated in the presence
(imposing the appropriate boundary conditions for) of strings, $Z_{\rm
  string}$ is the partition function for the trivial (constant) vacuum sector
of the model (and then Eq. (\ref{Fs}) is actuals the free energy difference
between the string and trivial vacuum sectors).  $\beta$, as always, is the
inverse of the temperature (we use throughout this work, unless explicitly
noted, with the natural units $\hbar,c,k_B=1$) and $L$ is the size of the
system.

The difficult with the approach given by (\ref{Fs}), which becomes evident
from Eq. (\ref{eq:5}) when we are dealing with nonconstant background fields,
is the nonlocal terms that appears in higher order perturbation terms when
expanding the effective action (in this case, the effective action for the
string background configurations). The only terms amenable of analysis are the
one-loop leading order terms.  Analogous approach based on the semiclassical
method, is the direct evaluation of the {\it classical} partition function
taking into account the string degrees of freedom, as performed by the authors
in Ref. \cite{copeland} and where the contribution and interpretation of the
phase transition based on the picture of string condensation is analyzed using
known statistical physics results.

Another approach that has been used is to define a field creation operator for
vortex-string excitations and then work directly with the correlation
functions in terms of these operators. This is the approach for instance taken
in Refs. \cite{MMR,marino}. However, also in this approach the evaluation of
correlation functions already at tree-level order is involved and results are
lacking beyond that order (though in the first reference of \cite{MMR} results
for the asymptotic behavior of the two-point correlation function for vortex
operators were obtained at one-loop order, but only for the $2+1$ dimensions
case).

\section{The Dual Action for Vortex-Strings}
\label{dualaction}

Let us start by writing the partition function for the Abelian Higgs model
(\ref{lagr}), which, in Euclidean space-time is given by

\begin{equation}
Z_\beta =\int \mathcal{D}A\mathcal{D}\phi \mathcal{D}\phi^*
\exp\left\{ -\int_0^\beta d \tau \int d^3 x
{\cal L}_E - S_{GF} \right\}  \;,
\label{ZAphi}
\end{equation}

\noindent
where in the above expression ${\cal L}_E$ denotes the Lagrangian density
(\ref{lagr}) in Euclidian space-time and $S_{GF}$ is some appropriate
gauge-fixing and ghost term that must be added to the action to perform the
functional integral over the relevant degrees of freedom.  A dual action to
the original one is obtained from (\ref{ZAphi}) by appropriately performing
Hubbard-Stratonovich transformations on the original field in such a way to
become explicitly the strings degrees of freedom, like in the form of Eq.
(\ref{chiparts}) and (\ref{vort2}).  {}For that, we first write the complex
Higgs field $\phi$ in the polar parameterization form $\phi =\rho e^{i\chi
}/\sqrt{2}$. Then, the scalar phase field $\chi $ is split in its regular and
singular terms, like in Eq. (\ref{chiparts}).

Lets for now, for convenience, omit the gauge fixing term $S_{GF}$ in Eq.
(\ref{ZAphi}) and re-introduce it again in the final transformed action.
{}Following e.g. the procedure of Refs.
\cite{klee,orland,chernodub,antonov1,kleinert}, the functional integral over
$\chi $ in Eq. (\ref{ZAphi}) can then be rewritten as

%EQUATION IN BOTH COLUMNS
\begin{widetext}
%\begin{equation}
%\end{equation}
\begin{eqnarray}
\lefteqn{\int \mathcal{D}\chi \,\exp \left[ -\int d^4x\frac 12\rho ^2\left(
\partial _\mu \chi +eA_\mu \right) ^2\right] }  \nonumber \\
&=&\int \mathcal{D}\chi _{\mathrm{sing}}\,\mathcal{D}\chi _{\mathrm{reg}}
\mathcal{D}C_\mu \left( \prod_x\rho ^{-4}\right) \,\exp \left\{ -\int
d^4x\left[ \frac 1{2\rho ^2}C_\mu ^2-iC_\mu \left( \partial _\mu
\chi _{\mathrm{reg}}\right) -iC_\mu \left( \partial _\mu \chi _{\mathrm{sing}
}+eA_\mu \right) \right] \right\}  \nonumber \\
&=&\int \mathcal{D}\chi _{\mathrm{sing}}\left( \prod_x\rho ^{-4}\right)
\mathcal{D}W_{\mu \nu }\,\exp \left\{ -\int d^4x\left[ \frac{\kappa ^2}{
2\rho ^2}V_\mu ^2+e\kappa A_\mu V_\mu +i\pi \kappa W_{\mu \nu }\omega _{\mu
\nu }\right] \right\} \;,
\label{dual2}
\end{eqnarray}
\end{widetext}

\noindent 
where we have performed the functional integral over $\chi _{\mathrm{reg}}$ in
the second line of Eq. (\ref{dual2}).  This gives a constraint on the
functional integral measure, $\delta (\partial _\mu C_\mu )$, which can be
represented in a unique way by expressing the $C_\mu $ in terms of an
antisymmetric field, $C_\mu = -i\frac \kappa 2\epsilon _{\mu \nu \lambda \rho
}\partial _\nu W_{\lambda \rho }\equiv \kappa V_\mu $, which then leads to the
last expression in Eq. (\ref{dual2}). $\kappa $ is some arbitrary parameter
with mass dimension and $\omega _{\mu \nu }$ is the vorticity, defined by Eq.
(\ref{vort2}) for the singular phase part of $\chi$.  Next, in order to
linearize the dependence on the gauge field in the action we introduce a new
antisymmetric tensor field $G_{\mu \nu }$ through the identity

\begin{eqnarray}
&&\exp \left( -\frac 14\int d^4x {}F_{\mu \nu }^2\right) \nonumber \\
&& =\int \mathcal{D}G_{\mu
\nu }\,\exp \left[ \int d^4x\left( -\frac{\mu _W^2}4G_{\mu \nu }^2-\frac{\mu
_W}2\,\tilde{G}_{\mu \nu }F_{\mu \nu }\right) \right] \;,
\label{dual3}
\end{eqnarray}
with $\tilde{G}_{\mu\nu} \equiv
\frac{1}{2}\epsilon_{\mu\nu\lambda\rho}G_{\lambda\rho}$.

\noindent
Substituting Eqs. (\ref{dual2}) and (\ref{dual3}) back into the partition
function, we can immediately perform the functional integral over the $A_\mu $
field.  This leads to the constraint $\epsilon_{\mu \nu \alpha \beta
}\partial_\mu \left( G_{\alpha \beta }-W_{\alpha \beta }\right) =0$ can be
solved by setting $G_{\mu \nu }=W_{\mu \nu }-\frac 1{\mu _W}\left( \partial
  _\mu B_\nu -\partial _\nu B_\mu \right)$, where we defined, for convenience,
$e\kappa =\mu _W$ and $B_\mu$ is an arbitrary gauge field. Using these
expressions back in the partition function (and re-introducing the gauge
fixing term) we then finally obtain the result

\begin{eqnarray}
Z&=&\int \mathcal{D}W_{\mu \nu
}\mathcal{D}\chi_{\mathrm{sing}}\,\mathcal{D}B_\mu\,\mathcal{D}
\rho \,\left( \prod_x\rho ^{-3}\right) \nonumber \\
&\times& \exp\left\{ -S_{\rm
dual}\left[ W_{\mu \nu },B_\mu, \rho ,\chi_{\rm sing} \right]
-S_{GF}\right\}, 
\label{ZWBrho}
\end{eqnarray}

\noindent
where the dual action is given by

\begin{widetext}
\begin{equation}
S_{\rm dual}= \int d^4x\left[ \frac{\mu _W^2}{2e^2\rho ^2}V_\mu
^2+\frac 14\left( \mu _WW_{\mu \nu }-\partial _\mu B_\nu +\partial
_\nu B_\mu \right) ^2+ \frac 12\left( \partial _\mu \rho \right)
^2-\frac{m_\phi ^2}2\rho ^2+\frac \lambda {4!}\rho ^4+i\pi
\frac{\mu _W}eW_{\mu \nu }\omega _{\mu \nu }\right] \;.
\label{SWBrho}
\end{equation}
\end{widetext}

\noindent 
The model described by $S_{\rm dual}$ is completely equivalent to the original
Abelian Higgs model, in the polar representation obtained from Eq.
(\ref{ZAphi}) and so, any calculations done using (\ref{ZWBrho}) must lead to
the same results as those done with the original action. The advantage of this
dual formulation (\ref{SWBrho}) is that it explicitly exhibits the dependence
on the singular configuration of the Higgs field, making it appropriate to
study phase transitions driven by topological defects. At the same time it
also show, from the last term in Eq. (\ref{SWBrho}), that the vortex-string's
degrees of freedom coupled to the matter field through the antisymmetric (or
Kalb-Ramond) field.  Now, if we come to the part concerning the gauge fixing
term $S_{GF}$ in (\ref{ZWBrho}), we see from Eq. (\ref{SWBrho}) that the dual
action exhibits invariance under the double gauge transformation: the
hypergauge transformation

\begin{eqnarray}
\delta W_{\mu \nu }(x) &=&\partial_\mu \xi_\nu (x)-\partial_\nu \xi_\mu
(x)\;,  \nonumber \\
\delta B_\mu &=&\mu_W\xi_\mu (x)\;,
\label{gauge1}
\end{eqnarray}
and the usual gauge transformation

\begin{equation}
\delta B_\mu =\partial_\mu \theta (x)\;,  
\label{gauge2}
\end{equation}

\noindent 
where $\xi_\mu (x)$ and $\theta (x)$ are arbitrary vector and scalar
functions, respectively. Choosing $\xi_\mu =B_\mu $ in the first
transformation is equivalent to fix the gauge through the condition $B_\mu =0$
\cite{orland} and this is equivalent to choose the unitary gauge in Eq.
(\ref{ZWBrho}).  The complete form for the gauge fixing action accounting for
the gauge invariances (\ref{gauge1}) and (\ref{gauge2}) was obtained in Ref.
\cite{dualpaper}, which, besides an overall normalization factor independent
of the action fields (and the background Higgs field) give for the quantum
partition function the complete result \cite{dualpaper}

\begin{widetext}
\begin{eqnarray}
Z &=&N\int \mathcal{D}W_{\mu \nu }\,\mathcal{D}\rho \,\mathcal{D}B_\mu \,
\mathcal{D}\overline{\eta }\,\mathcal{D}\eta \,\exp \left\{ -\int d^4x\left[
\frac{{\mu _W}^2}{2e^2\rho ^2}V_\mu ^2+\frac 14\left( \mu _WW_{\mu \nu
}-\partial _\mu B_\nu +\partial _\nu B_\mu \right) ^2\right. \right.
\nonumber \\
&+&\left. \left. \frac 12\left( \partial _\mu \rho \right) ^2-
\frac{m_\phi ^2}2\rho ^2+\frac \lambda {4!}\rho ^4-
\overline{\eta }\rho ^{-3}\eta -\frac
1{2\theta }\left( \partial ^\mu W_{\mu \nu }\right) ^2+\frac u{2\theta }
\mu_W W_{\mu \nu }\left( \partial ^\mu B^\nu -\partial ^\nu B^\mu \right)
+\frac{1}{2\xi }\left( \partial _\mu B^\mu \right) ^2\right] \right\} .
\label{Zgaugefix}
\end{eqnarray}
\end{widetext}

\noindent 
where $\overline{\eta }$, $\eta $ are the ghost fields used to exponentiate
the Jacobian $\rho ^{-3}$ in the functional integration measure in Eq.
(\ref{ZWBrho}) and $\theta ,u$ and $\xi $ are gauge parameters.

\section{The Effective Potential for local Vortex-Strings Averaged Fields}
\label{sec3}

Let us turn now to the study of the problem of vortex-strings condensation
during a phase transition. Thus, to proceed further with the evaluation of the
string contribution to the partition function we introduce a (nonlocal) field
associated to the string.  Quantizing the vortex--strings as nonlocal objects
and associating to them a wave function $\Psi [C]$, a functional field, where
$C$ is the closed vortex--string curve in Euclidean space-time, and noting
that the interaction term of the vortex-string with the antisymmetric field in
Eq. (\ref{SWBrho}) is in the form of a current coupled to the antisymmetric
field, following Refs. \cite{seo-suga,kawai} we can define the string action
term in the form

\begin{equation}
S_{\mathrm{string}}(\Psi [C],W_{\mu \nu })=\oint_Cdx_\nu \left[
|D_{\sigma^{\mu \nu }}\Psi [C]|^2-M_0^4|\Psi [C]|^2\right] \;,
\label{Sstring}
\end{equation}

\noindent
where $D_{\sigma^{\mu \nu }}$ is a covariant derivative term defined by
\cite{nambu2}

\begin{equation}
D_{\sigma ^{\mu \nu }}(x)=\frac \delta {\delta \sigma^{\mu \nu }(x)}-
i\frac{2\pi \mu _W}{e} W_{\mu \nu }(x)\;,
\label{covar}
\end{equation}

\noindent
where $\delta \sigma^{\mu\nu }(x)$ is to be considered as an infinitesimal
rectangular deformation on the string's worldsheet.  It can be easily checked
that Eq. (\ref{Sstring}) is invariant under the combined gauge invariances
(\ref{gauge1}) and (\ref{Psigauge}) if the hypergauge transformation
(\ref{gauge1}) is now supplemented by the vortex--string field transformation

\begin{equation}
\Psi [C]\to \exp \left[ -i\frac{2\pi \mu_W}e\oint dx^\mu \xi _\mu
(x)\right] \Psi [C]\;.
\label{Psigauge}
\end{equation}

\noindent
$M_0^4$ in Eq. (\ref{Sstring}) is a dynamical mass for the strings,

\begin{equation}
M_0^4\equiv \frac{1}{a^4} \left(e^{\tau_s a^2}-6\right)
\label{M0square}
\end{equation}

\noindent 
with $\tau_s$ is the string tension (the total energy per unit length of the
vortex-string) \cite{seo-suga,kawai}, which, in terms of the parameters of the
Abelian Higgs model, it is given by \cite{hindmarsh} $\tau_s = \pi \rho_c^2 \,
\epsilon (\lambda/e^2)$, where $\epsilon(\lambda/e^2)$ is a function that
increases monotonically with the ratio of coupling constants.  $a$ in Eq.
(\ref{M0square}) can approximately be given by the string typical radius can
be expressed as \cite{copeland})

\begin{eqnarray}
1/a  &\sim & m_\phi \left( 1-\frac{T^2}{T_c^2} \right)^{1/2} \;,
\label{aT}
\end{eqnarray}

\noindent
where $T_c$ is the mean-field critical temperature, $T_c = \sqrt{12
  m_\phi^2/(3 e^2 + 2\lambda/3)}$ \cite{dolan}.  $\rho_c$, the Higgs vacuum
expectation value, can likewise be expressed as

\begin{eqnarray}
\rho _c &\simeq &\sqrt{\frac{6m_\phi ^2}\lambda }
\left( 1-\frac{T^2}{{T_c}^2}\right) ^{1/2}\;.
\label{rhocT}
\end{eqnarray}

By defining a local string field as \cite{seo-suga}

\begin{equation}
\hat{\psi}_C\equiv 4\left( \frac{2\pi }e\right) ^2\sum_{C_{x,t}}
\frac{1}{a^3l}\left| \Psi [C]\right| ^2,
\label{local psic}
\end{equation}

\noindent 
where $l$ is the length of a curve $C$, and $C_{x,t}$ represents a curve
passing through a point $x$ in a fixed direction $t$. The vacuum expectation
value of $\hat{\psi}_C$ is denoted by $\psi _C$ and represents the sum of
existence probabilities of vortices in $C_{x,t}$. In terms of $\hat{\psi}_C$,
it can be shown that the contribution of the vortices to the quantum partition
function, indicated by the last term in Eq. (\ref{SWBrho}) and the integration
over $\chi_{\mathrm{sing}}$, can be written as \cite{seo-suga}

\begin{equation}
\int \mathcal{D}\Psi [C]\mathcal{D}\Psi ^{*}[C] e^{ -\int d^4x\left[
\frac 14\left( \frac e{2\pi }\right) ^2M_0^4\hat{\psi}_C+\frac{\mu _W^2}{4}
W_{\mu \nu }^2\hat{\psi}_C\right] }.
\label{vortexcontr}
\end{equation}

Eq. (\ref{vortexcontr}) implies, together with Eq. (\ref{SWBrho}), that an
immediate consequence of $\psi_C\neq 0$ is the increase of the $W_{\mu \nu}$
mass. This is directly associated with a shift in the mass of the original
gauge field in the broken phase, $M_A=e\rho_c$, as $M_A^2\rightarrow
M_A^2(1+\psi_C)$.  Since the field $\psi_C$, defined by Eq. (\ref{local
  psic}), works just like a local field for the vortex-strings, we are allowed
to define an effective potential for its vacuum expectation value $\psi_C$ in
just the same way as we do for a constant Higgs field. Since this
vortex-string field only couples directly to $W_{\mu \nu}$, at the one-loop
level the effective potential for $\psi_C$ will only involve internal
propagators of the antisymmetric tensor field. This effective potential, at
one-loop order and at $T=0$, was actually computed in Ref.  \cite{seo-suga} in
the Landau gauge for the antisymmetric tensor field propagator and it is given
by (in Euclidean momentum space and at finite temperatures)

\begin{eqnarray}
\lefteqn{ V_{\mathrm{eff}}^{\text{1-loop}}(\psi_C) = 
\frac{e^2}{4\pi^2 } M_0^4 \psi_C } \nonumber \\
&& +\frac{3}{2} \frac{1}{\beta} \sum_{n=-\infty}^{+\infty}
\int \frac{d^3 k}{(2\pi )^3}\ln \left[
\omega_n^2 + {\bf k}^2+M_A^2(1+\psi_C) \right]\;,
\label{VeffT0}
\end{eqnarray}

\noindent
where $\omega_n=2 \pi n/\beta$ are the Matsubara frequencies for bosons.  When
$\psi_C=0$, in the absence of string vacuum contributions to the partition
function, we re-obtain the standard result for the one-loop contribution to the
Higgs effective potential coming from the gauge field loops.  The sum over the
Matsubara frequencies in (\ref{VeffT0}) is easily performed \cite{dolan}.  We
can also work with the resulting expression by expanding it in the
high-temperature limit $M_A\sqrt{1+\psi_C}/T \ll 1$ and for $e^2/\lambda \ll
1$, which corresponds to deep in the second order regime of phase transition
for the scalar abelian Higgs model. This is analogous to the phenomenology of
the Landau-Ginzburg theory for superconductors, where the parameter
$(e^2/\lambda)^{-1}$ (also called the Ginzburg parameter), measuring the ratio
of the penetration depth and the coherent length, controls the regimes called
Type II and Type I superconductors.  In our case, the coherent length is
governed by $a\sim 1/M_H$, where $M_H$ is here the temperature dependent Higgs
mass, while the penetration depth is proportional to $1/M_A$, where $M_A$ is
the (temperature dependent) gauge field mass. This way we find a manageable
expression for the finite temperature effective potential given by
\cite{dualpaper}

\begin{eqnarray}
\lefteqn{ V_{\text{eff},\text{string}}^{(\beta)}(\psi_C) \simeq
\left[  \frac{e^2}{4\pi^2 } M_0^4 +
\frac{3 e^2 \rho_c^2}{16\pi ^2 a^2} +
\frac{e^2 \rho_c^2}{8}\, T^2\right] \psi_C } \nonumber \\
&& -\frac{e^3 \rho_c^3}{4\pi} 
\left( 1+\psi_C \right)^{3/2} T -
\frac{3 e^4 \rho_c^4 \ln \left[ 2/{a T} \right] }{32\pi^2}  
\psi_C^2 \;,
\label{Veff-highT3}
\end{eqnarray}

\noindent
where $M_0$, $a$ and $\rho_c$ are given by Eqs. (\ref{M0square}), (\ref{aT})
and (\ref{rhocT}).

We can then see that the quantum and thermal corrections in the effective
potential for strings, Eq. (\ref{Veff-highT3}), are naturally ordered in
powers of $\alpha =e^2/\lambda$.  Therefore, in the regime $\alpha \ll 1$ the
leading order correction to the tree-level potential in Eq.
(\ref{Veff-highT3}) is linear in $\psi_C$, while the second and the third
correction terms are ${\cal O}(\alpha^{3/2})$ and ${\cal O}(\alpha^2)$,
respectively.  Thus, the linear term in $\psi_C$ controls the transition in
the deep second order regime since the other terms are all subleading in
$\alpha$. Thus, near criticality, determined by some temperature $T_s$ where
the linear term in Eq. (\ref{Veff-highT3}) vanishes,
$V_{\text{eff},\text{string}}^{(\beta )}(\psi_C)\sim 0$ in the $\alpha \ll 1$
regime. $T_s$ is interpreted as the temperature of transition from the normal
vacuum to the state of condensed strings, is then determined by the
temperature where the linear term in $\psi_C$ in Eq. (\ref{Veff-highT3})
vanishes and it is found to be related to the mean field critical temperature,
for which the effective mass term of the Higgs field, obtained from
$V_{\mathrm{eff}}^{(\beta)} (\rho_c)$, vanishes. Using again Eqs.  (\ref{aT})
and (\ref{rhocT}), with the result $\tau_s a^2 \sim {\cal O}(1/\lambda)$ and
in the perturbative regime $e^2 \ll \lambda \ll 1$, after some
straightforward algebra, we find the relation

\begin{equation}
\frac{T_c-T_s}{T_c} \sim {\cal O} \left( \frac{e^{-1/\lambda}}{\lambda^{2}}
\right) \left[ 1+ {\cal O}(\alpha)\right]\;,
\label{critical shift}
\end{equation}

\noindent
with next order corrections to the critical temperatures difference being of
order ${\cal O}(\alpha)$. This result for $T_s$ allows us to identify it with
the Ginzburg temperature $T_G$ for which the contribution of the gauge field
fluctuations become important. These results are also found to be in agreement
with the calculations done by the authors in Ref. \cite{copeland}, who
analyzed an analogous problem using the partition function for strings
configurations, in the same regime of deep second order transition.

{}For the case where the gauge fluctuations are stronger, {\it i.e.}, for
$\alpha =e^2/\lambda \gtrsim 1$, the second term in Eq.\ (\ref{Veff-highT3})
of order $\alpha^{3/2}$ induces a cubic term $\rho_c^3$ to the effective
potential, favoring the appearance of a first order phase transition instead
of a second order one.  Here we see that the non-trivial vacuum $\psi_c\ne 0$
above the critical temperature $T_s$ enhance the first order phase transition
by an amount $(1+\psi_c)^{3/2}$. Hence, since $T_s\sim T_c$, we see that the
driven mechanism of the first order transition can be interpreted as a melting
of topological defects. This mechanism is very well known in condensed matter
physics \cite{kleinert} and always leads to a first order phase transition
(except in two dimensions).

\section{Conclusions}
\label{conclusions}

We have interpreted here the phase transition in the scalar Abelian Higgs
model as a process of condensation of vortex-strings condensation.  Our
analysis was based on a dual realization of the original model in such a way
to make explicitly the vortex-strings degrees of freedom of the nontrivial
vacuum of the model. This way, by constructing a field theory model for string
fields, the finite temperature effective potential for a local expectation
value for the string field was obtained. The transition temperature obtained
from this effective potential, the temperature of transition from the normal
vacuum to the state of condensed strings, was then obtained and identified
with the Ginzburg temperature for which gauge field fluctuations become
important.

Possible extensions of this work could, for example, to include magnetic
monopoles, like in the context of the compact Abelian Higgs model
\cite{projection}, in which case monopoles could be added as external fields
in the dual transformations. The study of finite temperature effects and
possible consequences for the confinement picture in the dual superconductor
model, should be possible in the context of the study performed in this work.

\acknowledgments

The authors would like to thank Conselho Nacional de Desenvolvimento
Cient\'{\i}fico e Tecnol\'{o}gico (CNPq-Brazil), {}Funda{\c {c}}{\~{a}}o de
Amparo {\`{a}} Pesquisa do Estado do Rio de Janeiro (FAPERJ) for the financial
support. R.O.R. would like to thank the organizers of the conference Infrared
QCD in Rio for the invitation to talk about this work at the conference.

%\subsection{}

% FIGURES

%\begin{figure}[htbp]
%\begin{center}
%\includegraphics[width=8cm]{}
%\caption{}
%\end{center}
%\end{figure}

%FIGURES IN BOTH COLLUMNS
%\begin{figure*}[htbp]
%\begin{center}
%\includegraphics[width=8cm]{}
%\caption{}
%\end{center}
%\end{figure*}

%EQUATION IN BOTH COLUMNS
%\begin{widetext}
%\begin{equation}
%\end{equation}
%\end{widetext}

\end{document}